\documentclass[11pt,a4paper]{article}
\pdfoutput=1
\usepackage{jheppub}
\usepackage{slashed}
\usepackage{multirow}
\usepackage{color}
\usepackage{mathtools}

\newcommand{\slv}{\raise.15ex\hbox{$/$}\kern-.53em\hbox{$v$}}
\newcommand{\slF}{\raise.15ex\hbox{$/$}\kern-.53em\hbox{$F$}}
\newcommand{\slL}{\raise.15ex\hbox{$/$}\kern-.53em\hbox{$L$}}
\newcommand{\slP}{\raise.15ex\hbox{$/$}\kern-.53em\hbox{$P$}}
\newcommand{\slp}{\raise.15ex\hbox{$/$}\kern-.53em\hbox{$p$}}
\newcommand{\slq}{\raise.15ex\hbox{$/$}\kern-.53em\hbox{$q$}}
\newcommand{\slR}{\raise.15ex\hbox{$/$}\kern-.53em\hbox{$R$}}
\newcommand{\slQ}{\raise.15ex\hbox{$/$}\kern-.53em\hbox{$Q$}}
\newcommand{\slK}{\raise.15ex\hbox{$/$}\kern-.53em\hbox{$K$}}
\newcommand{\slk}{\raise.15ex\hbox{$/$}\kern-.53em\hbox{$k$}}
\newcommand{\slD}{\raise.15ex\hbox{$/$}\kern-.53em\hbox{$D$}}
\newcommand{\slC}{\raise.15ex\hbox{$/$}\kern-.53em\hbox{$C$}}
\newcommand{\slA}{\raise.15ex\hbox{$/$}\kern-.53em\hbox{$A$}}
\newcommand{\slSigma}{\raise.15ex\hbox{$/$}\kern-.53em\hbox{$\Sigma$}}
\newcommand{\slpartial}{\raise.15ex\hbox{$/$}\kern-.53em\hbox{$\partial$}}
\newcommand{\slcalP}{\raise.15ex\hbox{$/$}\kern-.63em\hbox{$\cal P$}}

\def\bs{\boldsymbol}
\def\del{\partial}
\def\bdel{\boldsymbol \partial}

\def\p{{\boldsymbol p}}

\def\q{{\boldsymbol q}}

\def\k{{\boldsymbol k}}

\def\n{{\boldsymbol n}}
\def\x{{\boldsymbol x}}
\def\y{{\boldsymbol y}}

\def\r{{\boldsymbol r}}
\def\z{{\boldsymbol z}}

\def\bkappa{{\boldsymbol \kappa}}

\def\qqb{{q\bar q}}
\def\sM{\text{med}}

\def\Iqbqb{\mathcal{R}_{\bar q}}
\def\Iqq{\mathcal{R}_{ q}}
\def\Iqqb{\mathcal{J}}

\def\Rq{\mathcal{R}_{q}}
\def\Rqb{\mathcal{R}_{\bar q}}
\def\J{\mathcal{J}}

\newcommand{\beq}{\begin{eqnarray}}
\newcommand{\eeq}{\end{eqnarray}}
\newcommand{\be}{\begin{eqnarray*}}
\newcommand{\ee}{\end{eqnarray*}}

\newcommand{\nn}{\nonumber\\ }

\title{Jet coherence in QCD media: the antenna radiation spectrum}

\author[a]{Yacine Mehtar-Tani}

\author[b]{and Konrad Tywoniuk}

\affiliation[a]{Institut de Physique Th\'eorique, CEA Saclay, 
F-91191 Gif-sur-Yvette, France}

\affiliation[b]{
Department of Astronomy and Theoretical Physics,
Lund University,
S\"olvegatan 14A, 
SE-22 362 Lund, Sweden}

\emailAdd{yacine.mehtar-tani@cea.fr}
\emailAdd{konrad.tywoniuk@thep.lu.se}

\abstract{
We study the radiation of a highly energetic partonic antenna in a colored state traversing a dense QCD medium. Resumming multiple scatterings of all involved constituents with the medium we derive the general gluon spectrum which encompasses both longitudinal color coherence between scattering centers in the medium, responsible for the well known Landau-Pomeranchuk-Migdal (LPM) effect, and transverse color coherence between partons inside a jet, leading, in vacuum, to angular ordering of the parton shower. We discuss shortly the onset of transverse decoherence which is reached in opaque media. In this regime, the spectrum consists of independent radiation off the antenna constituents.}

\preprint{LU-TP 11-19}
\arxivnumber{1105.1346}

\begin{document}
\maketitle

\section{Introduction}

One of the key objectives of the heavy-ion program at the LHC is to investigate properties of the quark-gluon plasma (QGP) using hard probes. In particular, one addresses in-medium modifications of the fragmentation properties of final state energetic particles that depart from the well-known fragmentation pattern in vacuum, for instance in $e^+e^-$ annihilation, proton-proton collisions etc., where no dense medium is formed. These modifications are assumed to be sensitive to local medium properties, such as the density, as well as their spatiotemporal evolution. Indeed, strong medium effects are observed in heavy-ion collisions for both single-inclusive leading particle spectra \cite{Adler:2003au,Adams:2003kv,Aamodt:2010jd} and two-particle correlations \cite{Adams:2003im,Adler:2005ee}.
While such measurements have reached a high level of sophistication, shedding light on qualitative aspects of the quark-gluon plasma, studies of intrajet distributions in heavy ion collisions have recently been initiated both at RHIC \cite{Putschke:2008wn,Salur:2008hs,Lai:2009zq} and LHC \cite{Aad:2010bu,Chatrchyan:2011sx,Chatrchyan:2012gt} with many promising results and prospects for the future. 

The increased experimental capabilities at these high energies have also triggered several efforts to improve the theoretical understanding of gluon radiation in the presence of a colored medium. Until recently, only the leading order one-gluon medium-induced emission spectrum off a highly energetic quark or gluon, which will be denoted BDMPS-Z throughout, was known \cite{Baier:1996kr,Baier:1996sk,Zakharov:1996fv,Zakharov:1997uu}. Equivalent formulations were also derived in \cite{Wiedemann:1999fq,Wiedemann:2000za,Wiedemann:2000tf,Gyulassy:2000fs,Gyulassy:2000er,Arnold:2001ba,Arnold:2001ms,Arnold:2002ja,MehtarTani:2006xq}. This spectrum measures radiative parton energy loss in the QGP and accounts for momentum broadening of the radiated gluon which undergoes multiple scattering in the medium. The characteristic broadening of the transverse momenta of such gluons, arising due to coherence effects between medium rescatterings, sets an upper bound on the energy of the induced radiation which, nevertheless, can be quite sizable in relatively opaque media.

Regrettably, since the process under consideration does not deal with interference effects between emitters, see section~\ref{sec:statart}, the extension to multi-gluon emissions is bound to rely on {\it ad hoc} conjectures. In order to study the importance of these radiative interferences, lately the gluon emission spectrum off a time-like quark-antiquark ($\qqb$) antenna was calculated. In \cite{MehtarTani:2010ma,MehtarTani:2011gf} we considered an antenna traversing a relatively thin medium, i.e., assuming only one scattering in the medium background potential. In \cite{MehtarTani:2011tz}, on the other hand, we resummed multiple scatterings in the limit of soft gluon emission. The aim of the present work is to generalize the latter results to arbitrary number of rescatterings of the quark, antiquark and gluon, thus extending the validity of our previous findings to arbitrarily opaque media and up to large gluon energies. We also briefly discuss the main difference between the direct and interference contributions which relates to the physics of decoherence of QCD radiation, thus making contact with our previous work \cite{MehtarTani:2010ma,MehtarTani:2011tz}.


The key result of this work is the derivation of the interference spectrum, cf. eq.~\eqref{eq:interferencespec}, while a complete and detailed discussion of the emerging physical picture is presented in \cite{MehtarTani:2011gf,MehtarTani:2012cy}, see also \cite{CasalderreySolana:2011rz} for a complementary discussion. The paper is structured as follows. At the outset, in section~\ref{sec:statart}, we discuss briefly the known coherence phenomena relevant for high-energy physics and heavy-ion collisions. Then, in section~\ref{sec:amplitude}, we present and solve the classical Yang-Mills equations for the $\qqb$ antenna, thus obtaining a compact expression for the medium-induced gluon field. Medium averages for the total spectrum are described in section~\ref{sec:generalcs} where we also present the novel interference spectrum $\Iqqb$, given in eq.~\eqref{eq:interferencespec}, which encodes the new ingredients of transverse coherence in medium. The general properties of this spectrum are also outlined in brief. In particular, the emergence of strong color screening in relatively opaque media leads to a ``memory loss" effect, discussed in section~\ref{sec:discussion}. In this case the induced gluon spectrum becomes the incoherent superposition of radiation off the quark and the antiquark. Finally, we summarize and conclude in Section~\ref{sec:conclusions}.

\section{Color coherence phenomena in a few words}
\label{sec:statart}

Coherence phenomena in jet physics have been extensively studied in the framework of perturbative QCD since the early 80's (see, e.g., \cite{Dokshitzer:1991wu,Bassetto:1984ik,Bassetto:1982ma,Konishi:1979cb} and references therein). It has been shown that they substantially affect experimental observables. The effect that proved to be crucial dealt with jet evolution in vacuum: the so-called color coherence.  The cascading of a jet, initiated by a hard parton, occurs in a coherent manner. In other words, subsequent parton branchings of the shower are not independent but depend on the characteristics of the previous branching. More precisely, it was found that successive parton branchings are ordered in angles \cite{Mueller:1981ex,Ermolaev:1981cm}. This limits the phase space for soft emissions that tend to occur at large angles. Experimentally, this fact is manifested in the depletion of soft hadrons in the single-inclusive particle distribution, which exhibits  the so-called humpbacked plateau.

The feature of angular ordering reflects the transverse dynamics of the jet. As a simple example, let us consider the radiation pattern of a $\qqb$ antenna with opening angle $\theta_{\qqb}$ in a color singlet state.
As long as the transverse wave length of the radiated gluon, $\lambda_\perp \sim 1/k_\perp$, is smaller than the transverse dipole separation when the gluon is formed,  $r_\perp\sim t_{\text{form}}\theta_{\qqb}$ (the formation time being defined as $t_\text{form}\sim \omega /k_\perp^2$ where $\omega$ and $k_\perp$ are the gluon energy and transverse momentum, respectively), the gluon resolves the color structure of the pair and therefore is radiated either off the quark or the antiquark. In the opposite case the radiation is strongly suppressed since the gluon cannot resolve the color structure of the antenna. In terms of angles, this means that gluons emitted at angles $\theta>\theta_{\qqb}$ are suppressed. Clearly, this coherence phenomenon arises as long as two or more emitting particles are involved. Generally speaking, small-angle radiation which resolves the independent emitters is incoherent while large-angle gluons are sensitive to the total charge of the system and are therefore emitted coherently. In the following we shall refer to this color coherence phenomenon by ``transverse coherence".

In the case of in-medium radiation, another type of coherence arises which we, in contrast to the above, will call ``longitudinal coherence", namely, when the medium is dense enough an energetic particle can scatter coherently off several scattering centers along its trajectory during its formation time, $t_{\text{form}}\sim N_{\text{coh}}\lambda$ ($ \lambda$ being the mean free path in the medium and $N_\text{coh}$ the number of coherent scattering centers). It was realized a some time ago that this is analogous to the Landau-Pomeranchuk-Migdal (LPM) effect known from QED. In particular, the medium-induced radiative spectrum is suppressed by a factor $1/\sqrt{\omega}$,  as compared to incoherent radiative spectra induced by a single scattering, due to destructive interferences between the $N_{\text{coh}}$ scattering centers. The BDMPS-Z spectrum  \cite{Baier:1996kr,Baier:1996sk,Zakharov:1996fv,Zakharov:1997uu} mentioned above, which takes into account the non-Abelian LPM effect, has been the main ingredient in most studies of parton energy loss in the QGP. 

For the study of jet evolution in heavy-ion collisions it therefore seems imperative to incorporate both the transverse and longitudinal coherence phenomena. However, such a unified approach is missing to date. Only recently were both coherence effects considered together in the context of antenna radiation in medium \cite{MehtarTani:2010ma, MehtarTani:2011tz,MehtarTani:2011gf,MehtarTani:2012cy,CasalderreySolana:2011rz}.

Let us here merely point out a general analogy between the two phenomena. Note how the transverse and longitudinal wavelengths, $\lambda_\perp$ and $\lambda_\parallel \sim t_\text{form}$, each are sensitive to the number of emitters and the number of scattering centers, respectively, determining the breakdown of independent radiation (in the former case) or scattering (in the latter) and the onset of coherence in both cases. In this work, we shall address the issue of both transverse and longitudinal coherences in the same setup, which we argue is the building block of in-medium jet calculus.

\section{Emission amplitude from classical Yang-Mills equations}
\label{sec:amplitude}

As in our previous works \cite{MehtarTani:2010ma,MehtarTani:2011tz,MehtarTani:2011gf}, we shall proceed in the framework of the classical Yang-Mills (CYM) equations, applicable for soft gluon radiation off an energetic charge \cite{MehtarTani:2006xq}. First, let us recall the amplitude of emitting a gluon with momentum $k\equiv(\omega,\vec k)$, given by the standard reduction formula,
\beq
\label{eq:redform0}
{\cal M}_\lambda ^{a}({\vec k})=\lim_{k^2\to 0 } \int d^4x\, e^{ik\cdot x}\, \square_x A^{a}_\mu(x)  \epsilon^\mu_\lambda({\vec k}) \,,
\eeq
where $\epsilon^\mu_\lambda({\vec k})$ is the gluon polarization vector while $A^a_\mu$, the classical gauge field, is the solution of the CYM equations, 
\beq
[D_\mu,F^{\mu\nu}] = J^\nu, 
\eeq
with $D_\mu\equiv \del_\mu-ig A_\mu$ and $F_{\mu\nu}\equiv \del_\mu A_\nu-\del_\nu A_\mu-ig[A_\mu,A_\nu]$. The covariantly conserved current, i.e., $[D_\mu,J^\mu]=0$,  describes the projectiles.

We shall carry out our calculation in the light-cone gauge (LCG) $A^+=0$,\footnote{The light-cone decomposition of the 4-vector $x\equiv(x^0,x^1,x^2,x^3)$ is defined as $x\equiv(x^+,x^-,\x)$, where $x^\pm\equiv(x^0 \pm x^3)/\sqrt{2}$ and $\x = (x^1,x^2)$.} where only the transverse polarization contribute to the cross-section, and $\sum_{\lambda} \epsilon^i_\lambda(\epsilon_\lambda^j)^\ast=\delta^{ij}$, where $i(j)=1,2$. Then, the differential gluon radiation spectrum is given by
\beq
dN = \sum_{\lambda=1,2} |{\cal M}_\lambda ^{a}({\vec k})|^2 \, \frac{d^3k}{(2\pi)^3\,2k^+} \,,
\eeq
where the phase space is $d^3k\equiv d^2\k \,dk^+$.

Consider an energetic  $q \bar q$ pair with momenta $p\equiv(E, \vec p)$ and $\bar p\equiv( \bar E, \vec{\bar p})$, respectively, created in the splitting of a highly virtual photon or gluon generated in a hard process and moving in the $+z$ direction. In the infinite energy limit, or equivalently for soft gluon radiation, this virtual state has a life-time too short to be resolved by the emitted gluon. Indeed, for soft gluons the pair looks like if it was produced instantaneously at $t=0$. This property is the basis of soft-collinear factorization.

In the absence of the medium, the classical eikonalized current that describes the pair created at time $t_0=0$ reads $J^\mu_{(0)}=J^\mu_{q\,(0)}+J^\mu_{\bar q \,(0)}+J^\mu_{3\,(0)}$, 
where, e.g., the quark vacuum current reads
\beq
\label{eq:current-vacuum}
 J^{\mu,a}_{q \,(0)} = g\frac{p^\mu}{E}\,\delta^{(3)}({\vec x}-\frac{\vec p}{E}t)\,\Theta(t) \, Q_{q}^a \,.
\eeq
In momentum space the total current reads
\beq
 J^{\mu,a}_{(0)} (k)=-i g\left( \frac{p^\mu}{p\cdot k+i\epsilon}\ \, Q_{q}^a+ \frac{\bar p^\mu}{\bar p\cdot k+i\epsilon}\ \, Q_{\bar q}^a -\frac{p_3^\mu}{p_3\cdot k+i\epsilon}Q_3^a  \right),
\eeq
where $Q_q$ denotes the color charge vector of the quark (and, analogously, $Q_{\bar q}$ for the antiquark). The third component of the current is needed for charge conservation, such that $k\cdot J_{(0)}=0$ which leads to $Q_{q}+Q_{\bar q}=Q_{3}$, while momentum conservation implies ${\vec p}_3=-{\vec p}-{\vec{\bar p}}$. For a singlet antenna $Q_3=0$. In the case of a colored antenna, the third component of the current does not contribute in the frame where $p_3\approx (0,p_3^-,{\bs 0})$ because of the gauge choice.

Taking the square of the total color charge vector, it is therefore possible to obtain the scalar product of two different charges, which in our case are 
\beq
Q_q\cdot Q_{\bar q} = \begin{dcases} -C_F & \quad\text{for a singlet antenna }(\gamma^\ast\to \qqb) \\ -C_F+C_A/2 & \quad \text{for an octet antenna }(g^\ast\to\qqb)\,, \end{dcases}
\eeq
where $C_F\equiv (N_c^2-1)/2N_c$ and $C_A\equiv N_c$ are the fundamental and adjoint color charge-squares of SU(3). The above reasoning can be extended to arbitrary color configurations of the particles involved.\footnote{For example, the charge scalar product for a pure gluon antenna, i.e., $g^\ast\to g_1 g_2$, is simply $Q_{g_1}\cdot Q_{g_2} = -C_A/2$, while gluon radiation off a quark, $q^\ast\to q g$, yields $Q_q\cdot Q_g=-C_F/2$.}

We shall focus on the the region of small angles defined by: $p^+, \,\bar p^+\gg |\p|,\,|\bar \p|\gg k^+\gg |\k|$. This is suitable in intrajet physics where one deals mainly with collimated particles. The entire $\qqb$ pair and gluon system is then strongly collimated in the $+z$ direction \cite{MehtarTani:2006xq}. A systematic way to perform this limit is to boost the medium in the opposite direction. Since we are only interested in asymptotic states, i.e., in probing the field $A^a(x)$ at large times, $x^+\to\infty$, the amplitude \eqref{eq:redform0} can be rewritten in a more convenient form (cf. appendix \ref{sec:redform}), namely
\beq
\label{eq:redform2}
{\cal M}_\lambda ^{a}({\vec k})= -\int\limits_{x^+=+\infty} dx^-d^2\x\, e^{ik\cdot x}\, 2\del^+_x  {\bs A}^{a}(x) \cdot {\bs \epsilon}_\lambda(\vec k)  \,,
\eeq
for on-shell gluons, i.e., $k^2=0$. With the gauge choice above only the transverse component of the gauge field is dynamical. Its linear response to the medium interaction reads \cite{MehtarTani:2006xq}
\beq
\label{eq:field1}
\square A^i-2ig\left[A_\sM^-,\del^+ A^i\right] =-\frac{\del^i}{\del^+}J^++J^i \,,
\eeq
where the medium field $A^\mu_{\text{med}}$ only has a negative light-cone component which, in the limit considered above, is related to the medium color source density through the Poisson equation $-\bdel^2 A_{\text{med}}^-(x^+,\x)=\rho_{\text{med}}(x^+,\x)$ \cite{MehtarTani:2006xq}. In Fourier space it reads
\beq
A^-_\text{med}(q) = 2\pi\, \delta(q^+) \int_0^\infty dx^+ \mathcal{A}_\text{med}(x^+,\q) \,e^{i q^- x^+} \,.
\eeq
The current is found from the continuity relation, $\del_\mu J^\mu = ig [A^-_\sM,J^+]$, which can be solved iteratively \cite{MehtarTani:2011tz} with
\beq
J^{\mu}_{q(m)} = ig\frac{p^\mu}{p\cdot \del }~[A^-_\sM,J^+_{q(m-1)}] \,,
\eeq
where the subscript $m$ denotes the order of the expansion in the medium field. For $m>0$, in momentum space we get
\begin{align}
\label{eq:current}
J^{\mu,a}_{q(m)}(k) &= -(ig)^{m+1} \frac{p^\mu}{p\cdot k} \left[ \prod_{i=1}^{m}\int\frac{d^2\q_i}{(2\pi)^2} \int_0^{x^+_{i+1}} \!\!dx^+_i\right. \nn
& \qquad \times \left. e^{i \frac{\p\cdot\q_{i} }{p^+} x^+_i}T\cdot {\cal A}_\sM(x^+_i,\q_i) \right]^{ab} Q_q^b \ e^{i \frac{p\cdot k }{p^+} x^+_m},
\end{align}
where $x^+_{m+1}=L$ is the total medium length and $T$ are the generators of SU(3) in the adjoint representation, so that
\beq
[T \cdot {\cal A}_\sM(x^+_i,\q_i)]^{ab}Q^b = -i f^{abc} {\cal A}^c_\sM(x^+_i,\q_i)Q^b\,,
\eeq
where $f^{abc}$ are the SU(3) structure constants. Summing over the number of possible interactions, $J_q^{\mu,a} = \sum_{m=0}^\infty J_{q(m)}^{\mu,a}$, yields
\beq
\label{eq:current-sol}
J^{\mu,a}_{q}(k) =- ig\frac{p^\mu}{p\cdot k }\left[ \delta^{ab}+\int_0^{L} dx^+\, e^{i\frac{p\cdot k}{p^+} x^+} \del^- U_p^{ab}(x^+,0)\right] \, Q_q^b\,,
\eeq
where $U_p$ denotes the Wilson line in the adjoint representation, tracing the trajectory of the quark which is given by its momentum $p$. It is found from the general definition of the Wilson line in the adjoint representation, given by
\beq
\label{eq:WilsonLine}
U(x^+,0;[\r])\equiv{\cal P}_\xi\exp\left[ig\int _0^{x^+} \!\!d\xi\, T\cdot A^-_\sM\left(\xi,\r(\xi) \right)\right] \,,
\eeq
where ${\cal P}_\xi$ denotes path ordering along $\xi$ and $\r$ is defined by the trajectory of the probe by setting
\beq
\label{eq:WilsonLineQuark}
U_p(x^+,0) \equiv \left. U(x^+,0;[\r]) \right|_{\r(\xi) = \xi \,\p/p^+} \,.
\eeq
Note that color indices are  omitted when they are obvious to alleviate the notations. The general medium-modified current, given in eq.~\eqref{eq:current-sol}, was obtained for the first time in \cite{MehtarTani:2011tz}. In coordinate space, it simplifies to 
\beq
J^{\mu}_{q}(x)=U_p(x^+,0)\,J^{\mu}_{q(0)}(x),
\eeq
with the vacuum current defined in eq.~\eqref{eq:current-vacuum}. Note that $U_p(x^+,0)=U_p(L,0)$ for $x^+>L$. 
Returning presently to the calculation of the gauge field, the solution of eq. \eqref{eq:field1} takes  the following form
\beq
\label{eq:field-sol}
A_q^i(x)=\int d^4 y\; G(x,y)\, \widetilde J_q^i(y)  , 
\eeq
where the modified current reads 
\beq
 \widetilde J^i=-\frac{\del^i}{\del^+}J^++J^i \,,
\eeq
and the retarded Green's function is defined by 
\beq
\left(\square- 2\,ig\,T\cdot A^-_{\text{med}} \del^+\right)\, G(x,y)=\delta^{(4)}(x-y).
\eeq
Note that this Green's function is invariant under translations along the $x^-$ direction due to the fact that the medium field depends only on $x^+$ and $\x$. This translational  symmetry yields the conservation of the gluon energy $k^+$ while traversing the medium and holds as long as $k^+\gg |\k|$. This property allows us to introduce another useful Green's function 
\beq
{\cal G}(x^+,\x\,;\,y^+,\y|k^+)=\int^{+\infty}_{-\infty}dx^- e^{i(x-y)^- k^+} 2\del^+_xG(x,y) \,,
\eeq
which obeys the following Schr\"odinger-like equation
\beq
\label{eq:GSchrodingerEq}
\left(i\del^-+\frac{\bdel^2}{2k^+}+g\, T\cdot A^-_{\text{med}}\right)\,{\cal G}(x^+,\x\,;\,y^+,\y|k^+)=i\delta(x^+-y^+)\delta(\x-\y)\,.
\eeq
The solution to eq.~\eqref{eq:GSchrodingerEq} is well known and can be expressed in terms of a path integral in the transverse plane, leading to
\beq
\mathcal{G}\left(x^+,\x; y^+,\y| k^+ \right) = \int\mathcal{D}[ {\bs r} ]\, \exp\left[i\frac{k^+}{2}\int_{y^+}^{x^+} \!\!d\xi \, \dot{{\bs r}}^2(\xi) \right] U(x^+,y^+; [\r]) \,,
\eeq
where $U(x^+,y^+; [\r])$ is defined in eq.~\eqref{eq:WilsonLine} and the boundary conditions are $\r(y^+)=\y$ and $\r(x^+)=\x$. The Green's function $\mathcal{G}$ describes simultaneously the color rotation of the emitted gluon together with its Brownian motion in the transverse plane due to the interactions with the background field \cite{Wiedemann:1999fq,Wiedemann:2000ez,MehtarTani:2006xq}.

Inserting the field solution \eqref{eq:field-sol} into eq.~\eqref{eq:redform2}, we obtain the amplitude
\begin{align}
\label{eq:fieldeq-coordinate}
{\cal M}^a_{\lambda,q} ({\vec k}) &=  \int\limits_{x^+=+\infty} \!\!d^2 \x \,e^{ik^-x^+-i\k\cdot \x} \int dy^+ dy^- d^2\y \, e^{ik^+y^-}\nn
& \quad \times\,{\cal G}^{ab}(x^+,\x;y^+,\y|k^+)\, U^{bc}_p(y^+,0)\,{\bs \epsilon}_\lambda\cdot {\widetilde {\bs  J}}^c_{q(0)}(y)Ê\,,
\end{align}
where $k^-=\k^2/2k^+$ for the on-shell gluon. Simplifying further by integrating over $y^-$ and $\y$ by making use of the vacuum current given in eq.~\eqref{eq:current-vacuum},  the amplitude in eq.~\eqref{eq:fieldeq-coordinate} finally becomes
\begin{align}
\label{eq:fieldeq-compact}
{\cal M}_{\lambda,q} ^{a}({\vec k}) &= \frac{g}{k^+} \int\limits_{x^+=+\infty} \!\!d^2 \x \,e^{ik^-x^+-i\k\cdot \x} \int_0^{+\infty} dy^+   e^{i\frac{k^+p^-}{p^+} y^+}  \nn
&\quad \times{\bs \epsilon}_\lambda\cdot\left(i\bdel_y+ k^+ \n\,\right) \,{\cal G}^{ab}(x^+,\x\,;\,y^+,\y\,|k^+)\Big|_{\y=\n\,y^+}  \,U^{bc}_p(y^+,0) Q^c_q \,,
\end{align}
where we have introduced the dimensionless vector $\n=\p/p^+$ (equivalently, $\bar\n=\bar\p/\bar p^+$ for the antiquark). The corresponding amplitude for the antiquark, $\mathcal{M}^a_{\lambda,\bar q}(\vec k)$, is found by substituting $p$ with $ \bar p$ and $Q_q$ with $Q_{\bar q}$.

\begin{figure}
\centering
\begin{minipage}{5cm} 
\includegraphics[width=\textwidth]{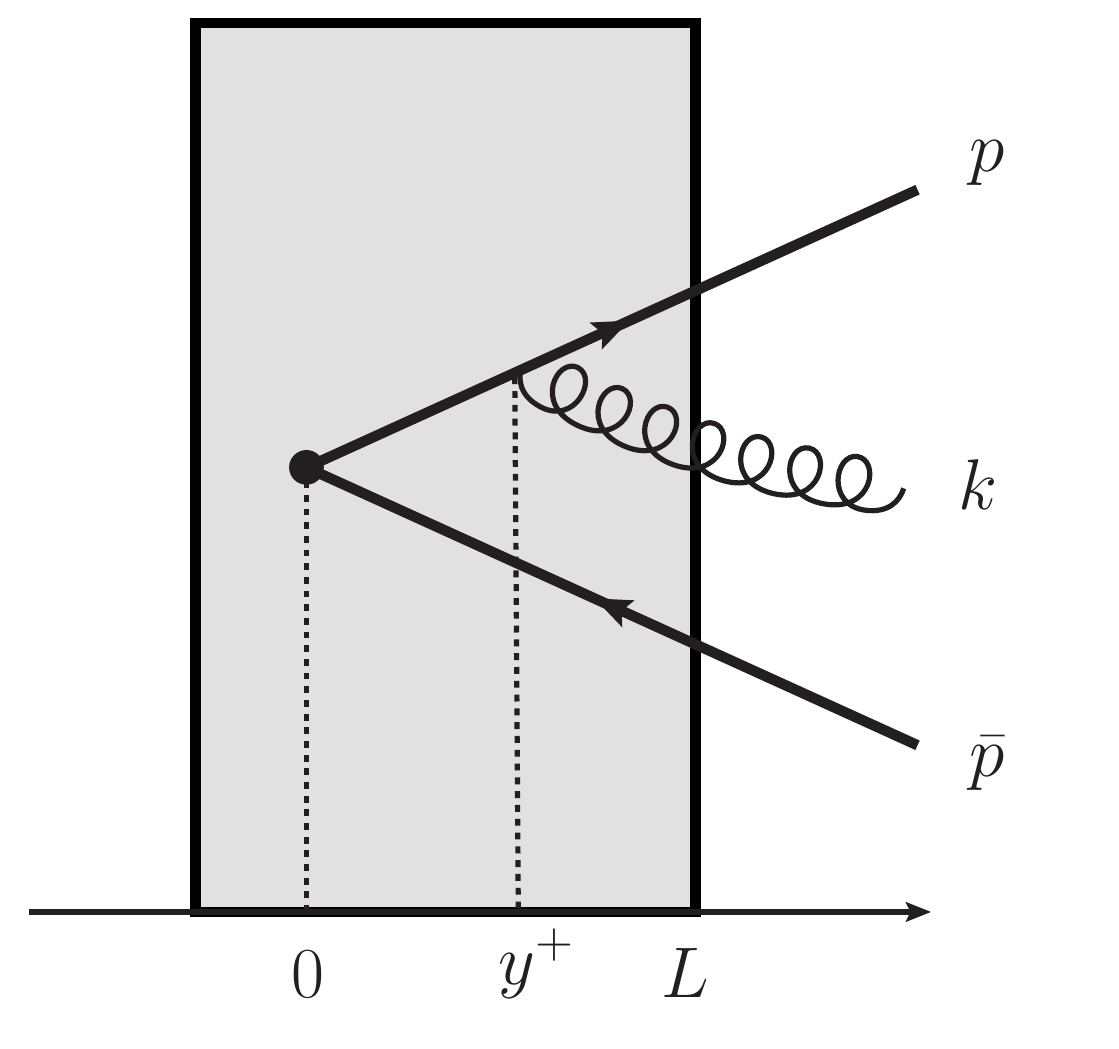}
 \end{minipage} 
 + sym.
\caption{The gluon radiation amplitude off an energetic  $\qqb$ pair created at $t=0$ in a medium of length $L$. The gluon is emitted at time $y^+$ (cf. eq.~\eqref{eq:fieldeq-compact}).}
\label{fig:AntennaMedium}
\end{figure}

Summarizing, the amplitude in eq.~\eqref{eq:fieldeq-compact} describes the propagation of the quark, with color charge $Q_q$, from the beginning of the medium up to a point $y^+$ where the gluon emission takes place, and the further propagation of the gluon through the medium out to the detector. The process is depicted in fig.~\ref{fig:AntennaMedium}. Note that $y^+>L$ implies gluon vacuum emission off an in-medium color rotated quark current.

\section{The antenna spectrum in the presence of a medium}
\label{sec:generalcs}

In the following, we assume a simple model for the medium, namely that it is made out of uncorrelated, static scattering centers (in the spirit of the Glauber picture). Then, we can treat the background field, $\mathcal{A}_\sM$, as a Gaussian white noise. The medium average of two fields can therefore be written as
\beq
\label{eq:medium-average}
\langle {\cal A}^a_\sM(x^+,\q) {\cal A}^{\ast b}_\sM(x'^+,\q')\rangle\equiv \delta^{ab}\,n(x^+) \,\delta(x^+-x'^+)(2\pi)^2 \,\delta^{(2)}(\q-\q'){\cal V}^2(\q)\,,
\eeq
where ${\cal V}(\q)$ is the medium interaction potential and $n(x^+)$ is the 3-dimensional density of scattering centers. In a static medium, $\mathcal{V}(\q)$ is usually chosen to be a Debye-screened Coulomb potential \cite{Wiedemann:1999fq,Wiedemann:2000ez,Wiedemann:2000za}.

\begin{figure}
\centering
\begin{tabular}{c c c  c}
\begin{minipage}{4.5cm} 
\includegraphics[width=\textwidth]{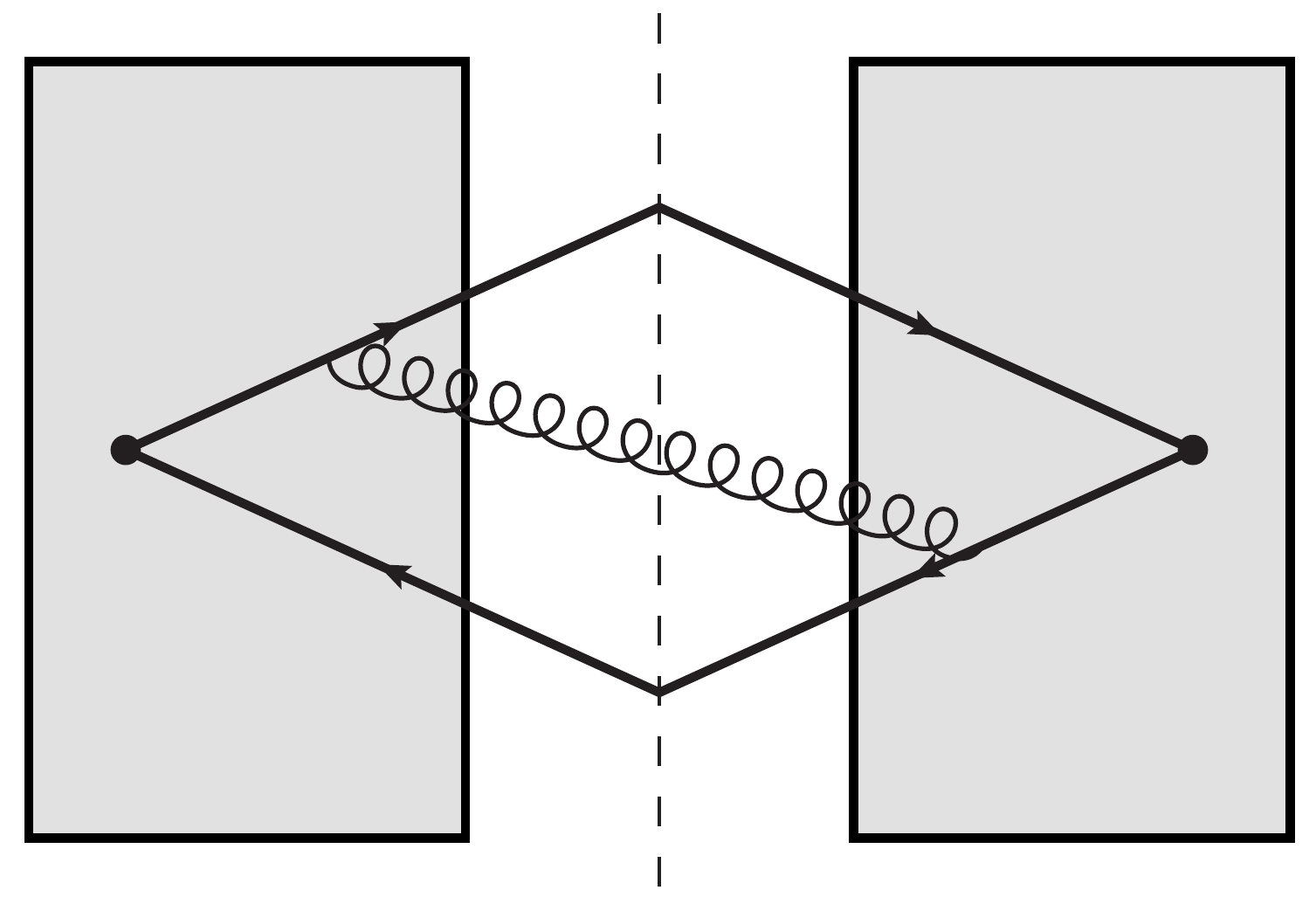}
 \end{minipage} &+&\begin{minipage}{4.5cm} 
\includegraphics[width=\textwidth]{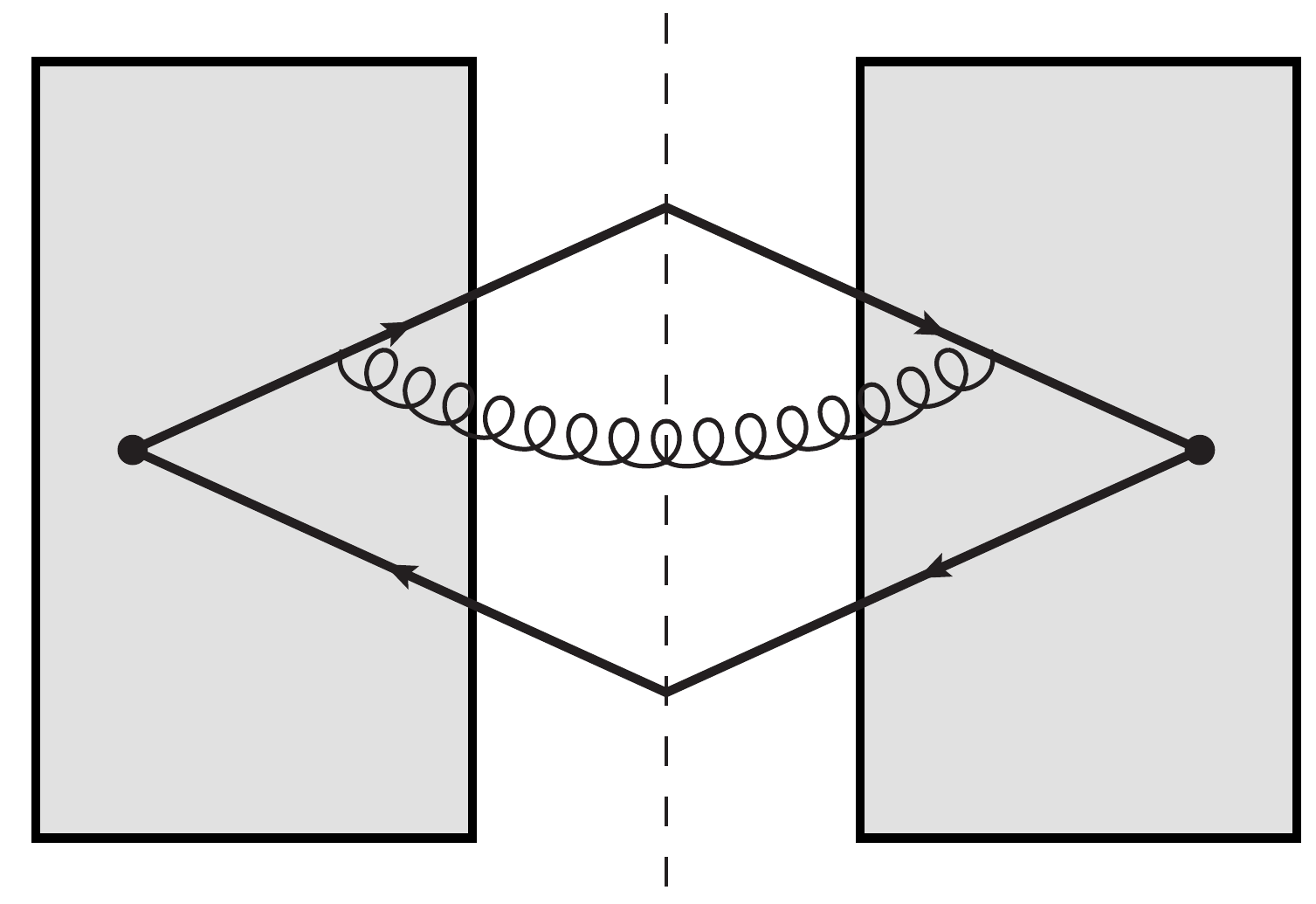}
 \end{minipage} &+sym.
\end{tabular}
\caption{Represention of the various contributions to the in-medium antenna spectrum. The first diagram stands for the interference part $\cal J$, while the second one stands for the BDMPS part ${\cal R}_q$. The rest symbolizes the symmetric configuration. }
\label{fig:CrossSection}
\end{figure}

Considering for the moment the most general case of a virtual gluon  splitting into a quark-antiquark pair, $ g^\ast\to q\bar q$, the spectrum can be written in the standard form \cite{Dokshitzer:1991wu} as 
\beq
\label{eq:decom-spect}
dN =\frac{\alpha_s}{(2\pi)^2} \left[C_F {\cal R}_{\text{sing}}+C_A\,\J\right] \, \frac{d^3k}{(k^+)^3} \,,
\eeq
where we have introduced the spectrum of a color-singlet antenna 
\beq
\label{eq:cohpart}
{\cal R}_{\text{sing}}=\Rq+ \Rqb-2\,\J,
\eeq
and
\begin{align} 
C_F\,\Rq &= (k^+)^2\langle|\mathcal{M}_q|^2\rangle \,,\\ 
C_F\,\Rqb &= (k^+)^2 \langle|\mathcal{M}_{\bar q}|^2\rangle \,\\ 
(-C_F+C_A/2)\,\J &=(k^+)^2\,\langle \text{Re}\, \mathcal{M}_q^\ast \mathcal{M}_{\bar q} \rangle \,,
\end{align}
represent the independent radiation components off the quark, the antiquark and the interferences, respectively. Here, $\langle \dotsb \rangle$ stand for medium averages as defined in eq.~\eqref{eq:medium-average}. Note that the first term in eq.~\eqref{eq:decom-spect} is proportional to the color charge of the quark/antiquark constituents whereas the second term is proportional to the total charge of the antenna. The total spectrum in eq.~\eqref{eq:decom-spect}, with its various components, is illustrated in fig.~\ref{fig:CrossSection}.

Because of the symmetries between the quark and the antiquark emission amplitudes we only need to evaluate the cross term, $\J$, represented in the left panel of fig.~\ref{fig:CrossSection}. Using eq.~\eqref{eq:fieldeq-compact} we obtain
\begin{align}
\J&=\,\text{Re}\left\{ \int_0^{+\infty} \!\!dy'^+ \int_0^{y'^+}\!\! dy^+ \int d^2\x \,\int d^2\x' e^{-i\k\cdot (\x-\x')+ik^+(\n^2y^+-\bar\n^2y'^+)/2}\right.\nn
& \quad \times\left(i\bdel_y+k^+\,\n\right) \cdot  \left(-i\bdel_{y'}+k^+\;\bar\n\right)\frac{1}{N_c^2-1}\langle {\text{{\bf Tr}}}\, {\cal G}(\infty,\x;y^+,\y|k^+)\, U_p(y^+,0)\nn
&\quad \times U^\dag_{\bar p}(y'^+,0){\cal G}^\dag(\infty,\x';y'^+,\y'|k^+)\, \rangle \Big\} + \text{sym.}\,,
\end{align}
where the gluon is emitted at $y^+$ in the amplitude and at $y'^+>y^+$ in the complex conjugate. The symmetric part, i.e., interchanging $q\leftrightarrow \bar q$ in all relevant quantities, accounts for the opposite $y^+$ ordering. Then, the medium average can be split into three pieces, namely
\begin{multline}
\frac{1}{N_c^2-1}\langle {\text{{\bf Tr}}}\, {\cal G}(\infty,\x;y^+,\y|k^+)\, U_p(y^+,0)U^\dag_{\bar p}(y'^+,0){\cal G}^\dag(\infty,\x';y'^+,\y'|k^+)=\\
\begin{split}
&\int d^2\z \frac{1}{N_c^2-1}\langle {\text{{\bf Tr}}}\, {\cal G}(\infty,\x;y'^+,\z|k^+){\cal G}^\dag(\infty,\x';y'^+,\y'|k^+)\rangle\\
&\qquad\times \frac{1}{N_c^2-1}\langle {\text{{\bf Tr}}}\, {\cal G}(y'^+,\z;y^+,\y |k^+)\, U^\dag_{\bar p}(y'^+,y^+)\rangle\\
&\qquad\times \frac{1}{N_c^2-1}\langle {\text{{\bf Tr}}}\, U_p(y^+,0)U^\dag_{\bar p}(y^+,0)\rangle \,,
\end{split}
\end{multline}
where we have used the following identity 
\beq
\mathcal{G}(\infty,\x;0,\y|k^+) = \int d^2\z \;\mathcal{G}(\infty,\x;z^+,\z|k^+) \mathcal{G}(z^+,\z;0,\y|k^+)\,.
\eeq
In the region from 0 to $y^+<y'^+$ the gluon is not produced yet, neither in the amplitude nor in the complex conjugate. The pure quark-antiquark interference is described by
\beq
\frac{1}{N_c^2-1}\langle {\text{{\bf Tr}}}\, U_p(y^+,0)U^\dag_{\bar p}(y^+,0)\rangle=1-\Delta_{\text{med}}(y^+,0),
\eeq
where $\Delta_{\text{med}}$ was defined in \cite{MehtarTani:2011tz} as the decoherence rate of the antenna,
\beq
\label{eq:deltamed}
\Delta_\text{med}(y^+,0) \equiv 1-\exp\left[-\frac{1}{2} \int_0^{y^+}\!\!\!\!d\xi\, n(\xi)\ \sigma(\delta \n \, \xi)\right]  \,,
\eeq
with $\delta \n = \n - \bar \n$. The modulus of this vector corresponds roughly to the opening angle of the pair, i.e., $|\delta\n|\equiv \sin \theta_{q\bar q}\sim \theta_{q\bar q}$. The decoherence rate depends on the dipole cross-section $\sigma$, which is given by
\beq
\label{eq:dipole-xsection}
\sigma(\delta \n\, \xi) = 2 \alpha_s C_A\int \!\! \frac{d^2\q}{(2\pi)^2} {\cal V}^2(\q)\left[1-\cos(\delta\n\cdot\q \,\xi)\right] \,.
\eeq
For consistency with previous calculations, we note that $n(\xi)\, \sigma(\r) \approx \frac{1}{2}\hat q(\xi) \,\r^2$ in the `harmonic oscillator' approximation \cite{Salgado:2003gb}, which is valid in the limit of multiple soft scatterings. Here, $\hat q$ is the medium transport coefficient  probing the accumulated transverse momentum squared per unit length.

From $y^+$ to $y'^+$, there is an interference between gluon and antiquark-medium interactions, which is described by
\begin{multline}
\frac{1}{N_c^2-1}\langle {\text{{\bf Tr}}}\, {\cal G}(y'^+,\z;y^+,\y |k^+)\, U^\dag_{\bar p}(y'^+,y^+) \rangle = \\
\begin{split}
& \exp\left\{ik^+\bar \n\cdot \left[\bar\z(y'^+)-\bar\y(y^+) \right]+i\frac{k^+}{2}\bar\n^2(y'^+-y^+)\right\} \\
& \times {\cal K}\left(y'^+,\bar\z(y'^+)\,;\,y^+, \bar\y(y^+) |k^+\right)\;,
\end{split}
\end{multline}
where $\bar\z(y'^+)=\z-\bar\n \,y'^+$ and $\bar\y(y^+)=\y-\bar\n\, y^+$. The gluon multiple scattering with the medium is taken into account by the path integral $\mathcal{K}$ \cite{Wiedemann:1999fq,Wiedemann:2000ez,Wiedemann:2000za}, given by
\beq
\mathcal{K}\left(y'^+,\z; y^+,\y| k^+ \right) = \int\mathcal{D} [{\bs r}] \, \exp\left[\int_{y^+}^{y'^+} \!\!d\xi \left(i\frac{k^+}{2} \dot{{\bs r}}^2(\xi) - \frac{1}{2} n(\xi) \sigma({\bs r}) \right)\right] \,,
\eeq
which describes the Brownian motion of the gluon the transverse plane from $\r(y^+)=\y$ to $\r(y'^+)=\z$. Finally, the medium average involving the gluon line element from $y'^+$ to $+\infty$ reads
\begin{multline}
\int d^2\x \int d^2\x' \frac{e^{-i\k\cdot(\x-\x')}}{N_c^2-1}\langle {\text{Tr}}\, {\cal G}(+\infty,\x;y'^+,\z|k^+){\cal G}^\dag(+\infty,\x';y'^+,\y'|k^+)\, \rangle=\\
\exp\left[-i\k\cdot(\z-\y')-\frac{1}{2}\int_{y'^+}^{+\infty} d\xi\, n(\xi)\,\sigma(\z-\y')\right] \,.
\end{multline}
For more details on medium averages see, for instance, \cite{Wiedemann:1999fq,MehtarTani:2006xq}. 

Putting everything together and after some straightforward algebra we obtain 
\begin{align}
\label{eq:interferencespec}
\J&= \text{Re}\left\{ \int_0^\infty dy'^+\int_0^{y'^+} dy^+ \big(1-\Delta_{\text{med}}(y^+,0) \big) \right. \nn
&  \quad \times \int d^2\z \,\exp\left[-i\bar\bkappa\cdot \z-\frac{1}{2}\int_{y'^+}^\infty d\xi\, n(\xi) \sigma(\z)+i\frac{k^+}{2}\delta\n^2 y^+\right] \nn
& \quad \times \left.  \left(\bdel_y-ik^+\,\delta\n\right)\cdot \bdel_z\,{\cal K}(y'^+,\z\,;\,y^+, \y\,|k^+)\right|_{\y=\delta\n y^+} \Bigg\}+\text{sym}. \,,
\end{align}
where $\bar\bkappa=\k-\bar x\bar\p$ (similarly, $\bkappa=\k-x\p$ ), and we have defined the light-cone momentum fractions $x=k^+/p^+$ and $\bar x = k^+/\bar p^+$. Equation \eqref{eq:interferencespec} is the main result of this paper. It describes the interference pattern between in-medium gluon radiation off a quark and an antiquark constituting a collimated antenna. 

Taking the limit $\delta \n \to 0$ in eq.~\eqref{eq:interferencespec} we find the spectrum off a single quark $\Iqq$, explicitly given by
\begin{align}
\label{eq:bdmpsspec}
\Rq&=\,2\,\text{Re}\left\{ \int_0^{\infty} \!\!dy'^+\int_0^{y'^+} \!\!dy^+ \int d^2\z \right.\exp\left[-i\bkappa\cdot \z-\frac{1}{2}\int_{y'^+}^\infty d\xi \, n(\xi) \sigma(\z)\right] \nn
& \quad \times  \bdel_y\cdot \bdel_z\,{\cal K}(y'^+,\z\,;\,y^+, \y ={\bs 0}\,|k^+)\Big\} \,,
\end{align}
where the factor 2 accounts for the opposite $y^+$ ordering. The antiquark spectrum $\Iqbqb$ is found analogously. This corresponds to the BDMPS-Z spectrum, in accordance with \cite{Wiedemann:1999fq,Wiedemann:2000ez,Wiedemann:2000za}.

\section{Decoherence in opaque media }
\label{sec:discussion}

The various parts of the total antenna spectrum \eqref{eq:decom-spect}, given in eq.~\eqref{eq:interferencespec} and eq.~\eqref{eq:bdmpsspec}, respectively, have many features in common. As mentioned before, both account for multiple scattering of all partons involved in the process---most importantly, the gluon---and therefore properly account for longitudinal coherence, which is reflected in the LPM suppression phenomenon. Apart from explicit dependence on the opening angle of the pair and the color charge, the interference terms of the antenna spectrum \eqref{eq:interferencespec} differs from the BDMPS-Z spectrum \eqref{eq:bdmpsspec} mainly by the appearance of factors that encompass the main elements of transverse coherence in medium. Postponing a exhaustive analysis of all the features of the gluon in-medium spectrum to a follow-up paper \cite{MehtarTani:2012cy}, presently we merely want to point out the most important of these. See also \cite{MehtarTani:2011tz} for a analysis valid for relatively dilute media.

The interferences, given in eq.~\eqref{eq:interferencespec}, are principally governed by the so-called decoherence factor $1-\Delta_{\text{med}}(y^+,0)$ \cite{MehtarTani:2011tz}, which is sensitive to the medium density and vanishes in the opaque limit. 
To estimate the relevant timescale for this depletion, let us presently assume that the constituents undergo multiple soft scatterings with a medium made out of independent scattering centers.\footnote{This implies that the range of interaction of these centers, or their screening length, should be much smaller than the mean free path $\lambda_\text{mfp}$.} Thus, in the ``harmonic oscillator" approximation \cite{Salgado:2003gb} the decoherence rate becomes
\beq
\Delta_{\text{med}}(t,0) =1-\exp\left(-\frac{1}{12}\hat q\, \theta_\qqb^2\, t^3 \right) \,.
\eeq
The decoherence rate tends to one exponentially controlled by the characteristic timescale for decoherence
\beq
\label{eq:tdecoh}
t_\text{d} \equiv \left( \hat q\, \theta_{\qqb}^2 \right)^{-1/3} \,. 
\eeq
Hence, when $\Delta_{\text{med}}(t_\text{d},0) \sim 1$ all interferences are suppressed and the pair decoheres.  It follows that for $t_\text{d}\ll L$, interferences are suppressed as compared to the incoherent contributions to the spectrum, as noted also in \cite{CasalderreySolana:2011rz}. This phenomenon can be understood as a screening effect in the presence of a dense medium \cite{MehtarTani:2011tz}. Not only does the quark (``daughter") loose sensitivity of the color charge of the antiquark (its ``sibling") but it also becomes oblivious to the color charge of the ``parent" gluon. This is reflected in the important fact that the spectrum in medium becomes independent of the total color charge of the antenna, see eq.~\eqref{eq:decom-spect}. It is nevertheless important to point out that the above considerations imply on the other hand that the interferences are instrumental at short timescales, $t < t_\text{d}$, and in particular in the regime where $t_\text{d} > L$. In this case, the interferences are weakly suppressed and decoherence is partial, see \cite{MehtarTani:2010ma,MehtarTani:2011gf,MehtarTani:2011tz}.

\section{Conclusions}
\label{sec:conclusions}

To summarize, we have computed the emission spectrum off a $\qqb$ antenna in both color singlet and octet representations traversing an arbitrary dense and colored medium. This generalizes our previous results in \cite{MehtarTani:2010ma}, which were valid only for dilute media, and in \cite{MehtarTani:2011tz}, where we only considered the strictly soft limit.

Our main result is the interference spectrum, given in eq.~\eqref{eq:interferencespec}, which encompasses two distinct types of QCD coherence. On one hand, the transverse momentum broadening of the gluon described by the path integral $\mathcal{K}$ is a manifestation of the longitudinal coherence which results in LPM suppression. On the other hand, due to the proper treatment of the transverse dynamics of the pair, $\Iqqb$ additionally contains interference between emissions off two different projectiles. The latter, radiative interferences, are crucial for building up the vacuum cascade and for accounting in a proper way for QCD coherence. Our  results thus establish generically how these effects are modified in the presence of an arbitrarily dense medium. We have argued that this leads to an exceptionally simple picture: the increase of the medium density leads to total decoherence of the pair. In effect, for dense enough media the emitters radiate independently of each other and, furthermore, they ``forget" about the existence of the total charge of their parent. 

The main result of this paper, eq.~\eqref{eq:interferencespec}, is quite general and a more detailed analysis of the antenna spectrum as a function of the medium characteristics goes beyond the scope of this study. The latter are fully contained in the two-point correlator eq.~\eqref{eq:medium-average}. For `dilute' media, where one is sensitive to rare medium interactions, a full analysis of the in-medium spectrum was already presented in \cite{MehtarTani:2011gf}. For a general discussion applicable for multiple soft scattering with the medium, where we can employ the ``harmonic oscillator" approximation as already mention in sec.~\ref{sec:discussion}, we refer the reader to \cite{MehtarTani:2012cy}, on the other hand. Finally, it is worth pointing that eq.~\eqref{eq:interferencespec} also includes the virtuality-driven spectrum that is still present even for vanishing medium characteristics, see also \cite{MehtarTani:2012cy} for a further discussion.

\section*{Acknowledgements}

We would like to thank Carlos Salgado and Nestor Armesto for stimulating discussions. This work is supported by Ministerio de Ciencia e Innovaci\'on of Spain; by Xunta de Galicia (Conseller\'{\i}a de Educaci\'on and Conseller\'\i a de Innovaci\'on e Industria -- Programa Incite); by the Spanish Consolider-Ingenio 2010 Programme CPAN; by the European Commission and in part by the Swedish Research Council (contract number 621-2010-3326).

\appendix
\section{The reduction formula eq.~\eqref{eq:redform2}}
\label{sec:redform}
In this appendix we shall see how we obtain the reduction formula in eq.~\eqref{eq:redform2}.  To alleviate the notation let us consider a generic field $A(x)$. Then, the amplitude for real gluon production is given by the LSZ reduction formula, 
\beq
\label{eq:redform1}
{\cal M}({\vec k})=\lim_{k^2\to 0 } \int d^4x\, e^{ik\cdot x}\, \square_x A(x) \,,
\eeq
where $x^+$ is evaluated up to some finite value which should then be sent to infinity. Explicitly, 
\begin{align}
{\cal M}({\vec k})&=\lim_{k^2\to 0 }\lim_{z^+\to +\infty}\int^{z^+} d^4x\, e^{ik\cdot x}\,  (2\del^+\del^--\bdel^2)A(x)\nn
&=\lim_{k^2\to 0 }\lim_{z^+\to +\infty}\Bigg[ \;\int dx^-d^2\x\, e^{ik\cdot x}\, 2\del^+ A(x)\Big|_{x^+=z^+} \nn
&\qquad\qquad\qquad-ik^-\int^{z^+} dx^+d^2\x \, e^{ik\cdot x} A(x)\Big|_{x^-=+\infty} \nn
&\qquad\qquad\qquad -k^2 \int^{z^+} d^4x\, e^{ik\cdot x} A(x) \Bigg]. 
\end{align}
The second term vanishes since the field vanishes in the limit $x^-\to \infty$. The remaining two terms are sensitive to the order of the limits. For example, taking the $z^+$ limit first leads to the standard relation
\beq
{\cal M}({\vec k}) = \lim_{k^2\to0} -k^2 A(k) \,,
\eeq
where the field is defined in momentum space. On the other hand, enforcing the on-shell condition from the outset, we end up with

\beq
\label{eq:redform}
{\cal M}({\vec k})=\lim_{x^+\to +\infty} \int dx^-d^2\x\, e^{ik\cdot x}\, 2\del^+_x A(x)  \,,
\eeq
where all quantities are defined in coordinate space and we have specified the space-time point the derivative acts upon.

\bibliography{mybib}{}

\providecommand{\href}[2]{#2}\begingroup\raggedright\begin{thebibliography}{10}

\bibitem{Adler:2003au}
{\bf PHENIX} Collaboration, S.~S. Adler {\em et.~al.}, {\it {High P(T) Charged
  Hadron Suppression in Au + Au Collisions at S(Nn$ )^1$/2 = 200-Gev}},  {\em
  Phys. Rev.} {\bf C69} (2004) 034910,
  [\href{http://xxx.lanl.gov/abs/nucl-ex/0308006}{{\tt nucl-ex/0308006}}].

\bibitem{Adams:2003kv}
{\bf STAR} Collaboration, J.~Adams {\em et.~al.}, {\it {Transverse Momentum and
  Collision Energy Dependence of High P(T) Hadron Suppression in Au + Au
  Collisions at Ultrarelativistic Energies}},  {\em Phys. Rev. Lett.} {\bf 91}
  (2003) 172302, [\href{http://xxx.lanl.gov/abs/nucl-ex/0305015}{{\tt
  nucl-ex/0305015}}].

\bibitem{Aamodt:2010jd}
{\bf ALICE} Collaboration, K.~Aamodt and C.~A. Loizides, {\it {Suppression of
  Charged Particle Production at Large Transverse Momentum in Central Pb--Pb
  Collisions at $\sqrt{S_{_{Nn}}} = 2.76$ TeV}},  {\em Phys. Lett.} {\bf B696}
  (2011) 30--39, [\href{http://xxx.lanl.gov/abs/1012.1004}{{\tt
  arXiv:1012.1004}}].

\bibitem{Adams:2003im}
{\bf STAR} Collaboration, J.~Adams {\em et.~al.}, {\it {Evidence from D + Au
  Measurements for Final-State Suppression of High P(T) Hadrons in Au + Au
  Collisions at Rhic}},  {\em Phys. Rev. Lett.} {\bf 91} (2003) 072304,
  [\href{http://xxx.lanl.gov/abs/nucl-ex/0306024}{{\tt nucl-ex/0306024}}].

\bibitem{Adler:2005ee}
{\bf PHENIX} Collaboration, S.~S. Adler {\em et.~al.}, {\it {Dense-Medium
  Modifications to Jet-Induced Hadron Pair Distributions in Au+Au Collisions at
  S(Nn$ )^($1/2) = 200- GeV}},  {\em Phys. Rev. Lett.} {\bf 97} (2006) 052301,
  [\href{http://xxx.lanl.gov/abs/nucl-ex/0507004}{{\tt nucl-ex/0507004}}].

\bibitem{Putschke:2008wn}
{\bf STAR} Collaboration, J.~Putschke, {\it {First Fragmentation Function
  Measurements from Full Jet Reconstruction in Heavy-Ion Collisions at
  $\sqrt{S_{_{\rm Nn}}}=200$ GeV by Star}},  {\em Eur. Phys. J.} {\bf C61}
  (2009) 629--635, [\href{http://xxx.lanl.gov/abs/0809.1419}{{\tt
  arXiv:0809.1419}}].

\bibitem{Salur:2008hs}
{\bf STAR} Collaboration, S.~Salur, {\it {First Direct Measurement of Jets in
  $\sqrt{S_{Nn}}=200$ GeV Heavy Ion Collisions by Star}},  {\em Eur. Phys. J.}
  {\bf C61} (2009) 761--767, [\href{http://xxx.lanl.gov/abs/0809.1609}{{\tt
  arXiv:0809.1609}}].

\bibitem{Lai:2009zq}
{\bf PHENIX} Collaboration, Y.-S. Lai, {\it {Probing Medium-Induced Energy Loss
  with Direct Jet Reconstruction in P+P and Cu+Cu Collisions at Phenix}},  {\em
  Nucl. Phys.} {\bf A830} (2009) 251c--254c,
  [\href{http://xxx.lanl.gov/abs/0907.4725}{{\tt arXiv:0907.4725}}].

\bibitem{Aad:2010bu}
{\bf Atlas} Collaboration, G.~Aad {\em et.~al.}, {\it {Observation of a
  Centrality-Dependent Dijet Asymmetry in Lead-Lead Collisions at Sqrt(S(Nn))=
  2.76 TeV with the Atlas Detector at the Lhc}},  {\em Phys. Rev. Lett.} {\bf
  105} (2010) 252303, [\href{http://xxx.lanl.gov/abs/1011.6182}{{\tt
  arXiv:1011.6182}}].

\bibitem{Chatrchyan:2011sx}
{\bf CMS} Collaboration, S.~Chatrchyan {\em et.~al.}, {\it {Observation and
  Studies of Jet Quenching in Pbpb Collisions at Nucleon-Nucleon Center-Of-Mass
  Energy = 2.76 TeV}},  {\em Phys. Rev.} {\bf C84} (2011) 024906,
  [\href{http://xxx.lanl.gov/abs/1102.1957}{{\tt arXiv:1102.1957}}].

\bibitem{Chatrchyan:2012gt}
{\bf CMS} Collaboration, S.~Chatrchyan {\em et.~al.}, {\it {Studies of Jet
  Quenching Using Isolated-Photon+Jet Correlations in Pbpb and PP Collisions at
  Sqrt(S[Nn]) = 2.76 TeV}},  \href{http://xxx.lanl.gov/abs/1205.0206}{{\tt
  arXiv:1205.0206}}.

\bibitem{Baier:1996kr}
R.~Baier, Y.~L. Dokshitzer, A.~H. Mueller, S.~Peigne, and D.~Schiff, {\it
  {Radiative Energy Loss of High Energy Quarks and Gluons in a Finite-Volume
  Quark-Gluon Plasma}},  {\em Nucl. Phys.} {\bf B483} (1997) 291--320,
  [\href{http://xxx.lanl.gov/abs/hep-ph/9607355}{{\tt hep-ph/9607355}}].

\bibitem{Baier:1996sk}
R.~Baier, Y.~L. Dokshitzer, A.~H. Mueller, S.~Peigne, and D.~Schiff, {\it
  {Radiative Energy Loss and P(T)-Broadening of High Energy Partons in
  Nuclei}},  {\em Nucl. Phys.} {\bf B484} (1997) 265--282,
  [\href{http://xxx.lanl.gov/abs/hep-ph/9608322}{{\tt hep-ph/9608322}}].

\bibitem{Zakharov:1996fv}
B.~G. Zakharov, {\it {Fully Quantum Treatment of the Landau-Pomeranchuk-Migdal
  Effect in Qed and QCD}},  {\em JETP Lett.} {\bf 63} (1996) 952--957,
  [\href{http://xxx.lanl.gov/abs/hep-ph/9607440}{{\tt hep-ph/9607440}}].

\bibitem{Zakharov:1997uu}
B.~G. Zakharov, {\it {Radiative Energy Loss of High Energy Quarks in
  Finite-Size Nuclear Matter and Quark-Gluon Plasma}},  {\em JETP Lett.} {\bf
  65} (1997) 615--620, [\href{http://xxx.lanl.gov/abs/hep-ph/9704255}{{\tt
  hep-ph/9704255}}].

\bibitem{Wiedemann:1999fq}
U.~A. Wiedemann and M.~Gyulassy, {\it {Transverse Momentum Dependence of the
  Landau-Pomeranchuk- Migdal Effect}},  {\em Nucl. Phys.} {\bf B560} (1999)
  345--382, [\href{http://xxx.lanl.gov/abs/hep-ph/9906257}{{\tt
  hep-ph/9906257}}].

\bibitem{Wiedemann:2000za}
U.~A. Wiedemann, {\it {Gluon Radiation Off Hard Quarks in a Nuclear
  Environment: Opacity Expansion}},  {\em Nucl. Phys.} {\bf B588} (2000)
  303--344, [\href{http://xxx.lanl.gov/abs/hep-ph/0005129}{{\tt
  hep-ph/0005129}}].

\bibitem{Wiedemann:2000tf}
U.~A. Wiedemann, {\it {Jet Quenching Versus Jet Enhancement: a Quantitative
  Study of the Bdmps-Z Gluon Radiation Spectrum}},  {\em Nucl. Phys.} {\bf
  A690} (2001) 731--751, [\href{http://xxx.lanl.gov/abs/hep-ph/0008241}{{\tt
  hep-ph/0008241}}].

\bibitem{Gyulassy:2000fs}
M.~Gyulassy, P.~Levai, and I.~Vitev, {\it {Non-Abelian Energy Loss at Finite
  Opacity}},  {\em Phys. Rev. Lett.} {\bf 85} (2000) 5535--5538,
  [\href{http://xxx.lanl.gov/abs/nucl-th/0005032}{{\tt nucl-th/0005032}}].

\bibitem{Gyulassy:2000er}
M.~Gyulassy, P.~Levai, and I.~Vitev, {\it {Reaction Operator Approach to
  Non-Abelian Energy Loss}},  {\em Nucl. Phys.} {\bf B594} (2001) 371--419,
  [\href{http://xxx.lanl.gov/abs/nucl-th/0006010}{{\tt nucl-th/0006010}}].

\bibitem{Arnold:2001ba}
P.~B. Arnold, G.~D. Moore, and L.~G. Yaffe, {\it {Photon Emission from
  Ultrarelativistic Plasmas}},  {\em JHEP} {\bf 11} (2001) 057,
  [\href{http://xxx.lanl.gov/abs/hep-ph/0109064}{{\tt hep-ph/0109064}}].

\bibitem{Arnold:2001ms}
P.~B. Arnold, G.~D. Moore, and L.~G. Yaffe, {\it {Photon Emission from Quark
  Gluon Plasma: Complete Leading Order Results}},  {\em JHEP} {\bf 12} (2001)
  009, [\href{http://xxx.lanl.gov/abs/hep-ph/0111107}{{\tt hep-ph/0111107}}].

\bibitem{Arnold:2002ja}
P.~B. Arnold, G.~D. Moore, and L.~G. Yaffe, {\it {Photon and Gluon Emission in
  Relativistic Plasmas}},  {\em JHEP} {\bf 06} (2002) 030,
  [\href{http://xxx.lanl.gov/abs/hep-ph/0204343}{{\tt hep-ph/0204343}}].

\bibitem{MehtarTani:2006xq}
Y.~Mehtar-Tani, {\it {Relating the Description of Gluon Production in Pa
  Collisions and Parton Energy Loss in Aa Collisions}},  {\em Phys. Rev.} {\bf
  C75} (2007) 034908, [\href{http://xxx.lanl.gov/abs/hep-ph/0606236}{{\tt
  hep-ph/0606236}}].

\bibitem{MehtarTani:2010ma}
Y.~Mehtar-Tani, C.~A. Salgado, and K.~Tywoniuk, {\it {Antiangular Ordering of
  Gluon Radiation in QCD Media}},  {\em Phys. Rev. Lett.} {\bf 106} (2011)
  122002, [\href{http://xxx.lanl.gov/abs/1009.2965}{{\tt arXiv:1009.2965}}].

\bibitem{MehtarTani:2011gf}
Y.~Mehtar-Tani, C.~A. Salgado, and K.~Tywoniuk, {\it {The Radiation Pattern of
  a QCD Antenna in a Dilute Medium}},  {\em JHEP} {\bf 04} (2012) 064,
  [\href{http://xxx.lanl.gov/abs/1112.5031}{{\tt arXiv:1112.5031}}].

\bibitem{MehtarTani:2011tz}
Y.~Mehtar-Tani, C.~A. Salgado, and K.~Tywoniuk, {\it {Jets in QCD Media: from
  Color Coherence to Decoherence}},  {\em Phys. Lett.} {\bf B707} (2012)
  156--159, [\href{http://xxx.lanl.gov/abs/1102.4317}{{\tt arXiv:1102.4317}}].

\bibitem{MehtarTani:2012cy}
Y.~Mehtar-Tani, C.~A. Salgado, and K.~Tywoniuk, {\it {The radiation pattern of
  a QCD antenna in a dense medium}},
  \href{http://xxx.lanl.gov/abs/1205.5739}{{\tt arXiv:1205.5739}}.

\bibitem{CasalderreySolana:2011rz}
J.~Casalderrey-Solana and E.~Iancu, {\it {Interference Effects in
  Medium-Induced Gluon Radiation}},  {\em JHEP} {\bf 08} (2011) 015,
  [\href{http://xxx.lanl.gov/abs/1105.1760}{{\tt arXiv:1105.1760}}].

\bibitem{Dokshitzer:1991wu}
Y.~L. Dokshitzer, V.~A. Khoze, A.~H. Mueller, and S.~I. Troian, {\it {Basics of
  Perturbative QCD}}, . Gif-sur-Yvette, France: Ed. Frontieres (1991) 274 p.
  (Basics of).

\bibitem{Bassetto:1984ik}
A.~Bassetto, M.~Ciafaloni, and G.~Marchesini, {\it {Jet Structure and Infrared
  Sensitive Quantities in Perturbative QCD}},  {\em Phys. Rept.} {\bf 100}
  (1983) 201--272.

\bibitem{Bassetto:1982ma}
A.~Bassetto, M.~Ciafaloni, G.~Marchesini, and A.~H. Mueller, {\it {Jet
  Multiplicity and Soft Gluon Factorization}},  {\em Nucl. Phys.} {\bf B207}
  (1982) 189.

\bibitem{Konishi:1979cb}
K.~Konishi, A.~Ukawa, and G.~Veneziano, {\it {Jet Calculus: a Simple Algorithm
  for Resolving QCD Jets}},  {\em Nucl. Phys.} {\bf B157} (1979) 45--107.

\bibitem{Mueller:1981ex}
A.~H. Mueller, {\it {On the Multiplicity of Hadrons in QCD Jets}},  {\em Phys.
  Lett.} {\bf B104} (1981) 161--164.

\bibitem{Ermolaev:1981cm}
B.~I. Ermolaev and V.~S. Fadin, {\it {Log - Log Asymptotic Form of Exclusive
  Cross-Sections in Quantum Chromodynamics}},  {\em JETP Lett.} {\bf 33} (1981)
  269--272.

\bibitem{Wiedemann:2000ez}
U.~A. Wiedemann, {\it {Transverse Dynamics of Hard Partons in Nuclear Media and
  the QCD Dipole}},  {\em Nucl. Phys.} {\bf B582} (2000) 409--450,
  [\href{http://xxx.lanl.gov/abs/hep-ph/0003021}{{\tt hep-ph/0003021}}].

\bibitem{Salgado:2003gb}
C.~A. Salgado and U.~A. Wiedemann, {\it {Calculating Quenching Weights}},  {\em
  Phys. Rev.} {\bf D68} (2003) 014008,
  [\href{http://xxx.lanl.gov/abs/hep-ph/0302184}{{\tt hep-ph/0302184}}].

\end{thebibliography}\endgroup
\bibliographystyle{jhep}

\end{document}